\documentclass[prl,amsmath,amssymb]{revtex4}
\usepackage{graphicx,amsfonts,amssymb,amsmath,hyperref}
\usepackage{textcomp}
\usepackage{esint}
\newif\ifhyper
\hypertrue
\ifhyper
\hypersetup{
  citecolor = {green},
  colorlinks = {true}, % false
  urlcolor = {blue} % magenta
}

\def\q{{\bf q}}

 \date{\today}

\begin{document}

\title{Comment on ``Universal and Non-Universal Correction Terms of Bose Gases in Dilute Region: A Quantum Monte Carlo Study'' [J. Phys. Soc. Jpn. 91, 024001 (2022)]}

\author{Adam Ran\c con }
 
 \affiliation{Univ. Lille, CNRS UMR 8523 - PhLAM - Laboratoire des Lasers Atomes et Molécules, F-59000 Lille, France}

\maketitle

Recently, Masaki-Kato et al. have investigated the thermodynamics of a hardcore Bose gas on the cubic lattice in the dilute limit \cite{Masaki2022}. In this regime, the system is expected to be very-well described as a gas of bosons in the continuum with effective mass $m=\hbar^2/(2ta^2)$, with $t$ the hopping amplitude on the lattice and $a$ the lattice spacing,  and a contact interaction characterised by its s-wave scattering length $a_s$ \cite{Rancon2014b}. By comparing their numerical results obtained by quantum Monte-Carlo simulations to the results of Bogoliubov theory, they were able to estimate numerically the ratio $a_s/a=0.316(2)$, as well as the constant $c_3$ describing the non-universal next order correction in the expansion in $n a_s^3$, with $n$ the density.

The goals of this comment are two-fold. Firstly, we show that the s-wave scattering length of hardcore bosons on the cubic lattice can be computed analytically and reads \cite{Rancon2014arxiv}
 \begin{equation}
a_s/a=\frac{8 \sqrt{6} \pi ^2}{\Gamma \left(\frac{1}{24}\right) \Gamma
   \left(\frac{5}{24}\right) \Gamma \left(\frac{7}{24}\right) \Gamma
   \left(\frac{11}{24}\right)}\simeq 0.31487023\ldots
   \label{eq_as}
\end{equation}
where $\Gamma(z)$ is the Gamma function. This results can be obtained by computing the T-matrix of the Bose-Hubbard model in the limit of infinite interactions (implementing effectively the hardcore constraint). It is in agreement with the estimate obtained by Masaki-Kato et al.
Secondly, we show that the results obtained in \cite{Masaki2022} agree with first order spin-wave calculations that take the hardcore constraint into account exactly, without referring to the Bose-Hubbard model with finite interaction.

\paragraph{1) Analytic calculation of the s-wave scattering length-}
The s-wave scattering length is obtained from the low-energy limit $E\to0$ of the T-matrix $T(E)$, which is obtained by resumming the ladder diagrams describing the multiple interactions between two bosons \cite{Rancon2011a,Castin}. This generalizes directly to the case of bosons on a lattice with on-site interaction $U$, for which \cite{Rancon2011a}
 \begin{equation}
T(0)=\left(\frac1U+\int_\q\frac1{2\epsilon_\q}\right)^{-1},
\end{equation}
where $\epsilon_\q=2t(3-\cos(q_x)-\cos(q_y)-\cos(q_z))$ is the dispersion on the cubic lattice (such that $\epsilon_\q\simeq t \q^2$ at low momenta, from which we read the effective mass) and $\int_\q\equiv \int_{BZ}\frac{d^3q}{(2\pi)^3}$ with $BZ$ the first Brillouin zone $[-\pi,\pi[^3$. Here and thereafter, we use  the lattice spacing $a$ as units of length and $\hbar=1$.

The scattering length is defined by
 \begin{equation}
\frac{4\pi a_s}{m}=8\pi t a_s=T(0),
\end{equation}
and in the limit $U\to\infty$ this leads to
 \begin{equation}
a_s=\frac{1}{\int_\q\frac{4\pi t}{\epsilon_\q}}.
\label{eq_as2}
\end{equation}
Evaluating the integral numerically, one recovers the result quoted in the note added in proof of \cite{Masaki2022}, based on \cite{Castin}. However, it so happen that the integral 
 \begin{equation}
I_3=\int_\q\frac{1}{3-\cos(q_x)-\cos(q_y)-\cos(q_z)},
\end{equation}
known as one of Watson's triple integrals \cite{Watson}, can be performed analytically and gives the rather amazing result
 \begin{equation}
I_3=\frac{\sqrt{6}}{96\pi^3}\Gamma \left(\frac{1}{24}\right) \Gamma   \left(\frac{5}{24}\right) \Gamma \left(\frac{7}{24}\right) \Gamma
   \left(\frac{11}{24}\right),
\end{equation}
which allows us to obtain Eq.~\eqref{eq_as}. A similar analysis shows that on the square lattice, one finds the two-dimensional scattering length of hardcore bosons is $a_s/a=e^{-C}/(2\sqrt{2})\simeq 0.198506\ldots$ with $C$ the Euler constant \cite{Rancon2011a,Rancon2014b}. 

\paragraph{2) Spin-wave calculation of the ground-state energy of hardcore bosons-}

In the dilute limit, the thermodynamics of hardcore bosons can be recovered in at least two ways. Mapping hardcore bosons on the quantum spin-$1/2$ XY model in a transverse field, a perturbation theory in $1/S$ to lowest order (spin-wave or semi-classical approximation) corresponds to a Bogoliubov theory for magnons in the dilute limit. Careful comparison, between the $1/S$ expansion and quantum Monte-Carlo on the square lattice have shown that the spin-wave approximation is in very good agreement with the numerics already at this order \cite{Coletta2012}. Another possibility is to use a (quantum) Weiss theory for hardcore bosons, that treat the on-site contraint exactly and the hopping at mean-field. This can then be improved by including loop corrections, that can be organized within the framework of the (lattice) functional renormalization group \cite{Rancon2014b}. Both methods are equivalent, and the ground-state energy per unit-volume in the grand-canonical ensemble at lowest order (equivalent to Bogoliubov theory for dilute bosons) reads in d-dimensions \cite{Rancon2014b}
\begin{equation}
E_{GS}=-\frac{dt}{2}\left(1+\frac{\mu}{2dt}\right)^2+\frac{1}{2}\int_\q \left(E_\q-\epsilon_\q\right),
\label{eq_GS}
\end{equation}
where $E_\q=\sqrt{\epsilon_\q\left(\left(\frac{\mu}{2dt}\right)^2\epsilon_\q+2dt\left(1-\left(\frac{\mu}{2dt}\right)^2\right)\right)}$ is the Bogoliubov dispersion of the spin-waves. The first term corresponds to the mean-field approximation, while the second term is the $1/S$ correction (with $S=1/2$) coming from the fluctuations of the Bogoliubov excitations. This result is valid for $-2dt\leq \mu\leq 2dt$ and works very well compared to Monte-Carlo in this whole range in $d=2$ \cite{Coletta2012}. Since fluctuations are expected to be weaker in higher dimensions, this approximation is expected to be very good in dimension three.

In three dimensions, the dilute limit corresponds to $\mu=-6t +\delta\mu$, with $\delta\mu/t\ll 1$. Using Eq.~\eqref{eq_as2} and rewriting the energy as
\begin{equation}
E_{GS}=-\frac{\delta\mu^2}{16\pi t a_s}+\frac{1}{2}\int_\q \left(E_\q-\epsilon_\q-\frac{\delta\mu^2}{12t}+\frac{\delta\mu^2}{2\epsilon_\q}\right),
\end{equation}
a careful analysis shows that
\begin{equation}
\frac{1}{2}\int_\q \left(E_\q-\epsilon_\q-\frac{\delta\mu^2}{12t}+\frac{\delta\mu^2}{2\epsilon_\q}\right)=\frac{8t}{15\pi^2}\left(\frac{\delta\mu}{t}\right)^{5/2}+\mathcal O\left(\delta\mu^{3}\right),
\end{equation}
and we recover the ground-state energy of a dilute Bose gas in the grand-canonical ensemble, including the Lee-Huang-Yang correction
\begin{equation}
E_{GS}=-\frac{m\delta\mu^2}{8\pi a_s}\left(1-\frac{64}{15\pi}\sqrt{ma_s^2\delta\mu}+C_{\rm cor.}\right).
\end{equation}
Note that the correction $C_{\rm cor.}$, of order $ma_s^2\delta\mu$, is non-universal and comes in part from lattice effects, and we have estimated it numerically to be $C_{\rm cor.}=\alpha ma_s^2\delta\mu+o(ma_s^2\delta\mu)$, with a conservative estimate $\alpha\simeq 3.43(1)$. From this, we recover the energy per particles in the canonical ensemble,
\begin{equation}
E/N=\frac{2\pi a_s n}m\left(1+\frac{128}{15\pi}\sqrt{n a_s^3}+\frac{4(256-9\pi^2\alpha)}{9\pi}n a_s^3+\mathcal O(n a_s^3)\right).
\end{equation}
Note that correction term of order $n a_s^3$ explicitly written here is not the full contribution, as we expect the next order in the loop (or $1/S$) expansion to also contribute at this order (in addition to a universal term in $n a_s^3\log(n a_s^3)$). Masaki-Kato et al. have estimated the full contribution at order $n a_s^3$ to be $c_3=130(40)$, and with $\frac{4(256-9\pi^2\alpha)}{9\pi}\simeq -6.9(1)$ we conclude that the first order in $1/S$ contributes for a few percent to $c_3$ in the opposite direction than the next order.

In summary, we have shown that the results of Masaki-Kato et al. are in very good agreement with the analytical results from the spin-wave approximation describing a dilute gas hardcore bosons. The latter includes the Lee-Huang-Yang correction as well as some of the non-universal corrections at the next order in $n a_s^3$. We have also shown that the scattering-length $a_s$, Eq.~\eqref{eq_as}, naturally appears in the ground-state energy. This comment leaves open some interesting questions. On the simulation side, it would be interesting to see if using $a_s$ as an input, and not as a fitting parameter, would give a better estimate of $c_3$. It would also be interesting to compare the numerical results to the spin-wave result Eq.~\eqref{eq_GS} away from the dilute limit (i.e. in the range $|\mu|\leq 6t$), to see if it compares as favourably as in two-dimensions. On the theory side, it would be interesting to compute the next correction in $1/S^2$ to the energy (as has been done in two dimensions in \cite{Coletta2012}) to have a theoretical estimate of $c_3$ and see how the universal correction in $n a_s^3\log(n a_s^3)$ is recovered.

\bibliography{bibli,/home/adam/Dropbox/Articles/bibli_merged}

\end{document}